# Mechanochemical modeling of dynamic microtubule growth involving sheet-to-tube transition

Xiang-Ying Ji and Xi-Qiao Feng*

Institute of Biomechanics and Medical Engineering, Department of Engineering Mechanics, Tsinghua University, Beijing 100084, China

## ABSTRACT

Microtubule dynamics is largely influenced by nucleotide hydrolysis and the resultant tubulin configuration changes. The GTP cap model has been proposed to interpret the stabilizing mechanism of microtubule growth from the view of hydrolysis effects. Besides, the microtubule growth involves the closure of a curved sheet at its growing end. The curvature conversion also helps to stabilize the successive growth, and the curved sheet is referred to as the conformational cap. However, there still lacks theoretical investigation on the mechanical-chemical coupling growth process of microtubules. In this paper, we study the growth mechanisms of microtubules by using a coarse-grained molecular method. Firstly, the closure process involving a sheet-to-tube transition is simulated. The results verify the stabilizing effect of the sheet structure, and the minimum conformational cap length that can stabilize the growth is demonstrated to be two dimers. Then, we show that the conformational cap can function independently of the GTP cap, signifying the pivotal role of mechanical factors. Furthermore, based on our theoretical results, we describe a Tetris-like growth style of microtubules: the stochastic tubulin assembly is regulated by energy and harmonized with the seam zipping such that the sheet keeps a practically constant length during growth.

## INTRODUCTION

As the most rigid cytoskeletal element in cells, microtubules have a well-marked lattice structure, consisting of regularly arranged $\alpha$- and $\beta$-tubulin heterodimers (e.g., (1,2)). The dimers bind head-to-tail along their longitudinal direction into polar protofilaments, which, in turn, associate laterally in a staggered manner, rendering an elegant tubular structure. In spite of their mechanical firmness and lattice regularity, the dynamic and evolutionary attributes are intrinsic and essential for microtubules to fulfill their various significant functions in cell divisions and other intracellular biological processes (e.g., (2-4)).

---

*Correspondence: fengxq@tsinghua.edu.cn.

Microtubules suffer stochastic transitions between growing and shrinking. Such dynamic processes are highly coupled with the hydrolysis of nucleotides bonded on the assembled tubulins (5,6). Dimers in different nucleotide states assume different curvatures (7), and two distinct protofilament configurations, curved or straight, are resulted in, depending on whether GDP or GTP is bonded (8,9). The intrinsic bending characteristic of GDP-tubulins is incompatible with the canonical microtubule lattice (10). To elucidate the physical mechanisms by which a microtubule composed mainly of bending GDP-tubulins can still keep stable growth, the GTP cap model has been proposed (11,12). It hypothesizes that only when the rate of tubulin assembly exceeds the rate of GTP hydrolysis, can some newly-added GTP-tubulin layers maintain as a cap at the growing end of the microtubule. The GTP cap can sustain the uniform lattice and the microtubule growth, and its disappearance will cause the depolymerization. Despite the logical elegance of this cap model, yet there is a shortage of sufficient evidence for the existence of the GTP cap, and disputes exist about its size (13,14) and the inside GTP distributions (15,16).

Experimental observations also reveal that a microtubule end can assume far more colorful conformations than mere straight or curved (17). Typically, an open and outward-curved sheet is imaged at the growing end by cryoelectron microscopy (cryo-EM) (18,19). Such a sheet structure has been considered an interphase during microtubule growth (20-22). The growth is achieved by the sheet closure which involves a distinct transition of curvature, from longitudinal to lateral. The feasibility of this growth pathway has been experimentally validated by Nogales and coworkers (10). They showed that at low temperature, tubulins binding GMPCPP, a non-hydrolysable GTP analogue, can form ribbons, in which the protofilaments have a radial bend of about 5° between two adjacent dimers. As temperature rises, the ribbons directly convert into tubes. It is thus suggested that the GMPCPP ribbons structurally correspond to the curved sheets at the growing microtubule end (10). The sheet-to-tube growth mode can interpret, quite successfully, the formation of the seam in a microtubule. The seam, apparently being a linear lattice defect, may virtually offer important binding sites for associated proteins to help zip the microtubule (23).

Prominently, the above-described growth mode, which involves a conversion of end conformations, itself provides a stabilizing mechanism for microtubule growth (9,24). Due to this mechanism, which is referred to as the conformational cap model, the sheet is more stable than the zipped microtubule body (25). The sheet closure happen stochastically (9), and its complete closure into a blunt-ended tube would induce microtubule shrinkage (26). Despite this well-conceived essentiality of the sheet structure in microtubule growth, the detailed mechanisms remain vague.

Though the conformational cap and the GTP cap models offer two different stabilizing mechanisms, a microtubule does not face an either-or choice. Actually, the conformational cap does not contradict the GTP cap but provides further guarantee for a stable microtubule growth. Also, the conformational change contributes bonus energy accumulated in microtubule lattice in addition to the energy from GTP hydrolysis (27,28), i.e., the elastic energy caused by the curvature transition (25). However, the detailed relationship between these two caps and the relation between closure and hydrolysis remain unclear. Structure analysis reveals that only GTP-tubulins can form lateral contacts compatible with the microtubule lattice, and the hydrolysis is not necessary for sheet closure and microtubule growth (10). However, this conclusion only suggests that hydrolysis could happen after closure, but does not dictate a direct link between closure and hydrolysis. Further investigation is desired to determine whether closure triggers hydrolysis (18,29). On the other hand, if a sheet totally composed of GDP-tubulins can stabilize the growth phase (25,30), the essential role of pure mechanical factors in cell physiology is signified since the conformational cap works purely mechanically.

These previous experimental findings clearly demonstrate the intrinsic and strong mechanics-biochemistry coupling, which is prevalent and vital for the dynamic behavior of microtubules. A deeper understanding of the appealing microtubule characters calls for well-defined theoretical models. Much effort has already been directed towards the modeling of microtubule dynamics. The switch between growing and shrinking and the corresponding length variation have been pithily described by differential equations (e.g., (13,31,32)). A detailed probe of the evolutions of conformation and energy inevitably needs finer three-dimensional simulations. Some examples have been given by VanBuren et al. (33) and Molodtsov et al. (34), who successfully accounted for the strain energy changes induced by the association and disassociation of individual dimers. As yet, however, there is still a lack of theoretical investigation on the dynamics of the sheet-to-tube growth mode. The spatial energy distribution and variation, as well as the complete growth process, under mechanical–chemical coupling regulations have rarely been addressed.

Recently, we established a coarse-grained model for studying the macroscopic behavior of microtubules (35). Our simulation on the intricate sheet-ended microtubule conformation and the radial indentation technique demonstrated the efficacy of the model. The sheet structure is shown to be energetically stable. In this paper, we will employ this model to investigate the microtubule growth process which involves a sheet-to-tube transition and is mechanical-chemical coupled. We comprehensively calculate the potential energy by taking into account the intrinsic curvatures of both GTP- and GDP-tubulins. Events as subunit association and sheet closure are treated as stochastic processes regulated by the coupled changes of chemical association energy and mechanical potential energy. The influence of GTP hydrolysis is also taken into account. This study gives insight into the conformational cap hypothesis and the stabilizing mechanisms in the microtubule growth process.

## METHODS

### Coarse-grained model of microtubule

In our previous work, we have developed a model to simulate the dynamic behavior of microtubules by considering their structural complexity and mechanical-chemical coupling features. Fig. 1 shows our model, which take accounts of seven types of monomer interactions according to the actual biophysical condition. Each interaction potential is assumed to be a quadratic function regarding the corresponding degree of freedom of deformation. The definition of interactions and interaction constants are summarized in Table 1 and five of them are illustrated in the enlarged view in Fig. 1. This model was used to reveal the detailed conformation of a sheet-ended microtubule in which protofilaments are both bending and twisting, and to simulate the indentation test for measuring the mechanical properties of a microtubule (35). The results have demonstrated the efficacy of this model in representing a dynamic process involving structural and energy evolution. In the present paper, the model established in Ref. (35) will be further employed to study the growth process and stabilizing mechanism of microtubules.

To more exactly reflect the complete growth process of a microtubule, a sequence of events involving the addition of a new subunit, the closure of sheet, and the GTP hydrolysis will be incorporated in the simulations. We further develop the previous established algorithm (35) to achieve this goal. For a newly happening event, an iterative computation of monomer positions following the previous algorithm is continued until the total potential energy becomes stable. Then, a new event is allowed to occur, and the corresponding changes of conformation and energy are calculated in the same way. Thus a continuous dynamic process can be simulated.

## Free energy of association

For polymerizing polymers, the free energy of association can be divided into two additive parts (36,37). One is beneficial for association, including the free energy associated with the interfaces or bonds between subunits, called "bond energy". Let $G_{\text{long}}$ and $G_{\text{lat}}$ denote the bond energies for longitudinal and lateral associations, respectively. The other part is unfavorable for association, denoted by $G_s$. It is the free energy required to immobilize a subunit in the polymer, to which the most important contribution is the entropy lost due to association. Noncovalent bonds as electrostatic and hydrophobic interactions are a bit loose rather than rigid. Therefore, the subunits are not totally fixed but have some freedom to rotate and vibrate. Evidently, the loss of entropy in this case is less than that when the translational and rotational degrees of freedom are completely lost (38,39). Here, we take $G_{\text{long}} = -19$ $k_B$T/dimer, $G_{\text{lat}} = -4$ $k_B$T/dimer, and $G_s = 11$ $k_B$T/dimer. These values match the theoretical results of chemical reaction kinetics (40) and are verified by computer simulations (41).

Fig. 2 shows the free association energies for the assembly of a tubulin dimer at different sites. In the first case, a dimer filling into the gap between two long protofilaments (Fig. 2 *a*) will form two lateral bonds and a longitudinal bond with the pre-assembled subunits and, therefore, the corresponding association energy is calculated by $G = G_{\text{long}} + 2G_{\text{lat}} + G_s$. For the second case, a dimer associating at the side of a long protofilament (Fig. 2 *b*) gets a lateral bond and a longitudinal bond. Correspondingly, the association energy equals $G = G_{\text{long}} + G_{\text{lat}} + G_{\text{s}}$. In the third case, only a longitudinal bond can be formed when a dimer assembles at the crest of the microtubule (Fig. 2 *c*) and the association energy is $G = G_{\text{long}} + G_{\text{s}}$. The competition and balance of the association energy and the potential energy will dictate the assembly process.

## Integrated thermodynamic description of microtubule growth

A growing microtubule can experience three possible events, namely, the assembly of tubulins at the tip of each protofilament, the closure of the seam, and the hydrolysis for GTP-tubulins after polymerization. The probabilities for the occurrence of these dynamic events are regulated by the corresponding changes of energy in the microtubule.

The assembly of tubulins is dimer-based and can be described as a thermodynamic process. For a tubulin dimer to assemble into the microtubule, all the 13 protofilament tips are potential polymerization sites, corresponding to 13 possible equilibrium energy states for the growing microtubule. All the energies at these 13 states are calculated and compared with the initial energy, and the differences are denoted as $\Delta E_{\text{n}}$ ($n = 1-13$), respectively. $\Delta E_{\text{n}}$ includes the variations of the interaction energy $U$ and the free energy of association $G$ regarding the $n$ th protofilament, that is,

$$\Delta E_{\text{n}} = U_{\text{n}} - U_{\text{ini}} + G_{\text{n}}, \qquad (1)$$

where $U_{\text{ini}}$ and $U_{\text{n}}$ are the total potential energies before and after the assembly of a dimer, respectively. The probability for the occurrence of assembly at each tip, $p_{\text{n}}$, is assumed to be proportional with the absolute value of $\Delta E_{\text{n}}$.

To date, no direct experimental observation has been reported on the closure process of microtubules. Considering the lateral interaction is between monomers, we characterize the closure as monomer-based, i.e., the seam is zipped by the consecutive linking of monomer pairs. The closure process is simulated as follows. When a pair of monomers at the opposite edges of the sheet is about to bond, the relevant lateral interactions come into play and the whole conformation of the microtubule evolves to a new equilibrium.

Ambiguity also exists for the GTP hydrolysis process, particularly for the relationship between sheet closure and GTP hydrolysis, and the distribution of GTP-tubulins in the microtubule, adding difficulties to the modeling. In this paper, unless otherwise specified, we assume that the closed part has been totally hydrolyzed whereas the sheet has not. The hydrolysis for the whole helical turn at the sheet root is treated as a synchronous event with the closure of the monomer pair. This relation is irrespective of the causality of the two processes.

# RESULTS

## The sheet-to-tube transition process contributes stored energy to the microtubule lattice

It is widely accepted that the hydrolysis of GTP bonded on the added dimers helps the accumulation of energy constrained in the microtubule lattice. In addition, the conformational cap model proposes that the curvature transition during the sheet-to-tube transition would also contribute to the lattice energy (25). We simulate the closure process of a microtubule and analyze the variation of the potential energy by the presented model. The sheet structure composed of GTP-tubulins measures ten monomers in length at the start. Once a monomer pair closes, the hydrolysis and the associated conformation change of tubulins on the same helical turn happen simultaneously. We take this assumption as the standard model hypothesis and will further investigate different geometrical and chemical conditions of microtubules in the following subsections. Fig. 3 shows the evolution of the total potential energy and the seven components of interaction energy during a successive closure process of three monomers. The continuous sheet-to-tube transition process and the energy changes are given in Movie S1 in the Supporting Material. It is demonstrated that the closure is steadily propelled. Both the energy barrier and the energy difference between two subsequent equilibrium states are identical during the whole closure process. An activation energy of about $10^4$ $k_B T$ is needed for a single monomer pair to close. When a monomer pair has been zipped and the whole microtubule has evolved into stable, the accumulated energy amounts to about 2400 $k_B T$, which is about 10 time higher than the energy from hydrolysis (29,42).

## A mere conformational cap plays a stabilizing role in microtubule growth

Now we turn to consider the stabilizing effect of the sheet structure. Experimental works have already suggested that the blunt end is a metastable intermediate between shrinking and growing. In other words, a blunt end is not as stable as a sheet. Thus, a sheet-ended microtubule is in a stable growth state, and only when the whole sheet has closed into a tube, can the microtubule shrinkage happen (26). In the last subsection, it has been shown that the zipping of a monomer pair requires an activation energy much higher than the energy difference between the two equilibrium states. Therefore, an open sheet structure does help prevent the microtubule from shrinking by providing an energy barrier for the sheet-to-tube transition. The stabilizing effect of a conformational cap is thus supported by our simulations.

The relationship between the GTP cap and the conformational cap has long been speculated but remains unclear. Here, we test whether the conformational change caused by GTP hydrolysis will resist the sheet-to-tube process and thus inhibit the microtubule growth.

Two microtubule models in different nucleotide states of the sheet are tested and compared with the standard structure described in the last subsection. In the first model, we assume that the microtubule, including the extending sheet, is composed merely of GDP-tubulins. In the second model, the sheet is half-hydrolyzed and, in other words, the upper part of the sheet is composed of GTP-tubulins whereas the lower part is composed of GDP-tubulins. The two models have the same configuration. In the standard model and the half-hydrolyzed sheet model, both of which have GTP tubulins at the sheets, the GTP hydrolysis process of the lowest GTP-helical turn is accompanied by its closure. Fig. 4 *a* compares the energy evolutions during the closure process in the three microtubule models. It is found that the sheet-to-tube transition can continue even when the sheet has been totally hydrolyzed. For the three sheets with different GTP distributions, both the values of the activation energies and equilibrium energy differences are similar.

Nogales and the co-workers have stated that the bending of GDP-tubulins is incompatible with the formation of canonical lateral interaction in microtubules (10). Here, our modeling is based on the premise that the lateral interaction has been pre-defined except that along the seam. Though insufficient to completely testify the relation between the longitudinal bending conformation and the lateral interaction, our results demonstrate that the curvature transition and the sheet closure can be achieved by highly curved GDP-tubulins. This indicates that the conformational cap could be uncoupled with the GTP cap. Despite the likelihood that the substructure change in hydrolyzed tubulins would resist the formation of lateral interactions, at least from the view point of energy mechanism, hydrolysis is allowed to the tubulins on the sheet before closure and the microtubule growth will not be interfered by the nucleotide state. In addition, our simulations evidence the stabilizing role of a mere conformational mechanism in microtubule growth, and highlight the significance of mechanical factors in the microtubule behaviors.

# The intrinsic curvature of GTP-tubulins scarcely affects the function of conformational cap

Though straighter than GDP-tubulins (43), GTP-tubulins are not perfectly straight as widely assumed (44). A slight bend of about 5° was exhibited at the intra-dimer interface of each GTP-tubulin (45). Using the defined model, we now explore the influence of this intrinsic curvature of GTP-tubulins on the growth process of microtubules. The energy variations during the closure of GTP sheets with an intrinsic curvature of 5° (standard), 0° and 15° are compared in Fig. 4 *b*. It is seen that these three cases do not have distinct difference either in the activation energy or the equilibrium energy stepping. This means that neither the conformation nor the energy evolution during the sheet-to-tube transition is sensitive to the longitudinal curvature. The intrinsic curvature should be dictated mainly by some structural factors and seems to be not critical to the growth process and stabilizing mechanisms of microtubules. This result supports the recent investigation about the intrinsic bending of microtubule protofilaments by Grafmüller and Voth (46), who, through large-scale atomistic simulations, concluded that no observable difference exists between the mesoscopic properties of the intra-dimer and inter-dimer. The distinct curvature difference between polymerizing and depolymerizing protofilaments may majorly due to their lattice constraint.

## A conformational cap should include at least two tubulin dimer layers

The cap length has long been a controversial issue (16,47,48). In this subsection, we test the dependence of the energy barrier and equilibrium energy stepping during the zipping of a monomer pair on the sheet length from ten monomers to one. The results are shown in Fig. 5. Clearly, the energy barrier and the energy stepping for the sheets containing four to ten monomers in each protofilament along the length direction are almost identical, but they are much higher than those in shorter sheets. If the sheet is shorter than the length of two dimers, the activation energy to close the sheet drops distinctly and thus the whole open sheet would experience a swift and unstable transition into a tube. In this case, the sheet structure will lose its stabilizing effect and then depolymerization will occur.

Recent experimental and theoretical works about the size of GTP cap also conclude that an effective GTP cap should include at least two dimer layers (15,34). This minimum size of the GTP cap accords to that of the conformational cap estimated by our model. Such a consonance hints the potential direct relevance between closure and hydrolysis. As estimated by some researchers (29), hydrolysis can be catalyzed by the closure due to the energy accumulated.

## Microtubule growth plays Tetris

We have demonstrated that the open sheet structure at the growing end can stabilize the microtubule growth and have testified the sheet-to-tube growth style. However, if the closure speed is faster than the tubulin assembly rate at the tip, the sheet structure will vanish and depolymerization may be triggered at the blunt microtubule end. Therefore, a stable growth should rest on harmonized closure and polymerization, keeping the sheet length practically constant, of at least two dimer layers. In this case, the seam's zipping will experience a self-similar propagation, which conjures up the reverse process of a stable crack advancement.

Further, we extrapolate that the assembly and closure is regulated as Tetris. Tubulin assembly can happen at each protofilament tip. In view of the fact that the change of potential energy due to the assembly of a single tubulin is tiny and intrinsically fluctuates, the assembly site is majorly determined by the free energy of association. That is, the position A in Fig. 2 *a* is the most likely site to accommodate a tubulin dimer, whereas an assembly at position C in Fig. 2 *c* has the smallest possibility. This association process shares the idea of Kossel-Stranski model in the kinetic theory of crystal growth. When a top layer of the sheet has been stochastically filled in by the coming tubulins, the closure of a dimer-length will happen subsequently. Fig. 6 shows some snapshots of the "fill in–close up" process, and the dynamic process is shown in Movie S2. The sheet length is nearly constant so that the growth is steadily advanced; conversely, if the sheet is quickly closed into tube, the game will be soon over.

## DISCUSSIONS

### Physiological indications

*The actual sheet length*

The above calculations put forward the hypothesis that the sheet keeps a nearly constant length of more than two dimers during the sheet-to-tube growth process. However, the actual sheet length is yet unknown. We speculate that the sheet length may correlate with the nucleation process, and the length determined in the nucleation is sustained in the subsequent growth stage. It has been supposed that the microtubule nucleation template is likely to be some sheet structure composed of the laterally associated short protofilaments. The sheet closes to form an embryo of the tube (19,21). Thus the microtubule nucleation and assembly share the same basic mechanism (25). For a longitudinally curved sheet to transit its curvature and close into tube, it needs to overcome an energy barrier, which should be higher if more tubulins were zipped. The available energy that can be employed to overcome the barrier may determine the nucleated conformation and, in turn, the sheet length. It is inspiring that the sheet-to-tube nucleation pathway can be simulated under the same framework as presented in this paper. Our further study will focus on simulating the nucleation process to probe the critical nucleus, the structural templates, and the influence of nucleation on the subsequent growth.

Alternate likelihood is that the sheet length is relevant with the hunt of the microtubule ends for regulators of assembly such as microtubule-associated proteins (MAPs) and coming tubulins. As can be seen from the form-finding results of sheet-ended microtubules (35), the sheets with different lengths have different shapes: a shorter sheet has a rounded top edge and its lateral alignment is more compact. The sheet end offers special recognition sites for tubulins to assemble. The inter-protofilament interfaces may accommodate MAPs (45,49). Specifically, some plus-end-tracking proteins (+TIPs) bind the microtubule end with a higher affinity than the wall (50), and one of the proposed mechanisms is their recognition of a unique structural feature of the growing end (51,52). These effects should be of direct relevance with the space and structure of the edge line, the lateral gap between protofilaments, and, therefore the sheet length.

*Indicated roles of some +TIPs such as EB1/Mal3p*

Our simulations clearly reveal that there exists a steady energy barrier during the closure of a sheet, and the sheet-to-tube process needs to be activated. It has been supposed by experiments that a kind of microtubule +TIPs, EB1 in vertebrates and its homolog Mal3p in schizosaccharomyces pombe, can bind the seam. In the presence of this protein, microtubule growth is promoted (23). We guess that the binding of EB1/Mal3p can lower the activation energy required for a tubulin pair to bond and catalyze the sheet-to-tube process.

Besides the function of helping the seam to zip, +TIPs are speculated to have some other roles in promoting the microtubule growth. Firstly, +TIPs may facilitate the assembly of longer tubulin oligomers in addition to individual tubulin dimers by associating the dimers along its length in solution and adding the whole oligomer into the growing microtubule (53). Secondly, because of this template effect of +TIPs for oligomer assembly, the dimers are pre-straightened before adding into the microtubule, thus the whole microtubule structure is more firm and the closure is facilitated. These influences of assembly units and configurations on the global microtubule growth can also be conveniently investigated by this model.

### Mechanical influences of GTP hydrolysis

GTP hydrolysis is directly bound up with the microtubule dynamics. Tubulin dimers bind two GTP molecules. GTP at the N-site of *α*-tubulin is non-exchangeable, whereas GTP at the E-site of *β*-tubulin will hydrolyze into GDP after assembly (54). It is commonly agreed that the main body of a microtubule is made of GDP-tubulins, although the GTP distribution at the sheet structure at the growing end is yet unclear. However, some experiments suggest that the microtubule lattice also contains scattered GTP-tubulin remnants, meaning that the hydrolysis is sometimes incomplete during polymerization (55,56). Currently, the moment and condition for GTP to hydrolyze, as well as the influence of GTP hydrolysis on the properties and behavior of microtubules, is little known. In this paper, microtubule models with sheets in different nucleotide states have been compared. We have also tested the models in which the tubulins in the closed lattice are not in the same nucleotide state and, namely conformational state. No distinction is found with respect to the sheet-to-tube curvature conversion process, implying that the conformational change of tubulins resulted from GTP hydrolysis hardly interferes with the mechanic requirement of the global conformation evolution of a growing microtubule. The conformational change may mainly influence the depolymerization process. For example, it weakens the lateral interaction and facilitates the ram's horn-like peeling of protofilaments during shrinking, and the GTP remnants in the lattice could help rescue the microtubule from shortening (53,54).

## Model discussion

### Influences of the interaction definitions

In our model, three kinds of interactions that are not experimentally based, and their values are assumed as $k_{\text{lat}}^{\text{dihedral}} = k_{\text{lat}}^{\text{bend}} / 50$, $k_{\text{long}}^{\text{dihedral}} = k_{\text{long}}^{\text{bend}} / 50$, and $k_{\text{diag}} = k_{\text{long}}$ (35). Our previous work has demonstrated that the variations of the two dihedral angles have little influence on the total energy (35). The diagonal interaction majorly acts to restrain the fluctuations of tubulin positions and make the calculation converge faster. Here, we examine the influence of these values on the energy evolution during the closure process and validate our assumptions.

We vary each constant value independently by keeping the other two at their originally assumed values. Fig. 7 shows the results. No surprisingly, the 10-fold changes of $k_{lat}^{dihedral}$ and $k_{long}^{dihedral}$ do not make notable differences on the energy and conformation evolutions. With changing $k_{diagonal}$, the energy barriers alternate. A smaller $k_{diagonal}$-value results in a larger barrier, due to the increased flexibility of the model and the resultant larger tubulin displacements during the calculation process, but the stable state is soon found. Moreover, the equilibrium energy stepping remains the same and the energy barrier value is stable for each closure, so all conclusions in the paper can be validated.

*Limitations and further directions*

It is desired if a time factor is incorporated in the model. This requires the knowledge about the rate of assembly, closure and hydrolysis, by which a spatiotemporal evolution of conformation and energy could be elucidated. Besides, the bond rupture, namely, the depolymerization is not considered in the presented model. We hope that a systematic modeling of the integrated dynamic process of tubulin assembly, sheet closure, and protofilament peeling can be accomplished.

# CONCLUSION

In this study, we have simulated the chemical-mechanical coupled microtubule growth process which involves a sheet-to-tube transition. The conformation and energy evolution is exhibited and the stabilizing mechanism of the conformational cap is analyzed. We demonstrate that an effective conformation cap should comprise at least two dimer layers, and the cap length is maintained during a stable growth process by the harmonized tubulin assembly and sheet closure.

# SUPPORTING MATERIAL

Movie S1.
Movie S2.

This work was supported by the National Natural Science Foundation of China (Grant No. 10732050 to X.Q.F).

|   | **Interaction** | **Interaction potential** | **Interaction constant** | **Value** |
|---|---|---|---|---|
| 1 | Longitudinal tension or compression | $U_{\text{long}} = k_{\text{long}} \times d_{\text{long}}^2 / 2$ | $k_{\text{long}}$ | 3.0 nN/nm |
| 2 | Lateral tension or compression | $U_{\text{lat}} = k_{\text{lat}} \times d_{\text{lat}}^2 / 2$ | $k_{\text{lat}}$ | 14.0 nN/nm |
| 3 | Diagonal tension or compression | $U_{\text{diag}} = k_{\text{diag}} \times d_{\text{diag}}^2 / 2$ | $k_{\text{diag}}$ | 3.0 nN/nm |
| 4 | Longitudinal bending | $U_{\text{long}}^{\text{bend}} = k_{\text{long}}^{\text{bend}} \times q_{\text{long}}^2 / 2$ | $k_{\text{long}}^{\text{bend}}$ | 2.0 nN·nm |
| 5 | Lateral bending | $U_{\text{lat}}^{\text{bend}} = k_{\text{lat}}^{\text{bend}} \times q_{\text{lat}}^2 / 2$ | $k_{\text{lat}}^{\text{bend}}$ | 8.5 nN·nm |
| 6 | Longitudinal dihedral bending | $U_{\text{long}}^{\text{dihedral}} = k_{\text{long}}^{\text{dihedral}} \times y_{\text{long}}^2 / 2$ | $k_{\text{long}}^{\text{dihedral}}$ | 0.04 nN·nm |
| 7 | Lateral dihedral bending | $U_{\text{lat}}^{\text{dihedral}} = k_{\text{lat}}^{\text{dihedral}} \times y_{\text{lat}}^2 / 2$ | $k_{\text{lat}}^{\text{dihedral}}$ | 0.17 nN·nm |

Table 1 Interaction definitions in the model

# Figure Legends

FIGURE 1  Model of a sheet-ended microtubule. The enlarged view shows five of the defined interactions, including: (1) longitudinal tension or compression interaction, (2) lateral tension or compression interaction, (3) diagonal tension or compression interaction, (4) longitudinal bending interaction, and (5) lateral bending interaction. For the detailed mathematical depictions and the definitions of the longitudinal and lateral dihedral bending interactions, please refer to Ref. (35).

FIGURE 2  Sites for a coming tubulin dimer to assemble. (*a*) the dimer inserting into a gap, (*b*) the dimer associating a single-sided neighbor, and (*c*) the dimer falling upon the crest.

FIGURE 3  Potential energy evolution during the sheet-to-tube transition process. A continuous zipping of the seam counting three pairs of monomers is characterized. The consecutive closure happens when the microtubule is in an equilibrium conformation. The upper panel shows the evolution of the total potential energy, from which the energy barrier and energy difference between two equilibrium states are clearly detected. The lower panel exhibits the evolution of the seven energy components. Respecting the differences of orders of magnitude, a semilogarithmic coordinate is adopted.

FIGURE 4  Comparison of the energy barriers and energy differences during the sheet-to-tube transition under different sheet structures: (*a*) the sheets are in three different nucleotide states, (*b*) the intrinsic curvatures of GTP-tubulins are of three different values. In both panels, the red lines represent the result for the standard model shown in Fig. 3. For clarity, the three sets of data have been offset horizontally, but not vertically.

FIGURE 5  Energy barrier and energy stepping for a monomer pair closure of sheets of different lengths, varying from 1 to 10 monomers in the longitudinal direction. The ten microtubule models are composed of protofilaments of the same length. The total potential energies are at different levels since the length of closed parts of the ten models are different.

FIGURE 6  Coupled assembly and closure during microtubule growth. (*a*) Snapshots of the sheet structure. When a helical turn at the end has been fully filled by tubulins, the seam will be zipped a same length and the sheet will be closed up. Two subsequent closures are shown. (*b*) Schematic drawing of the sheet structure evolution exhibited in (*a*). Three stages colored black, blue and red in chronological order are involved. The short dashes at the bottom represent the root of the sheet, and the arrays of 13 short lines at the top represent the tubulin distributions at the tips of the 13 protofilaments. The first "fill in–close up" process is characterized in blue, and the second in red; the corresponding conformations are highlighted with the same colors in (*a*).

FIGURE 7  Influences of interaction constants on the energy barrier and energy difference

between two equilibrium states during closure. (*a–c*) Influences of the interaction constants of lateral dihedral, longitudinal dihedral, and diagonal tension or compression, respectively.

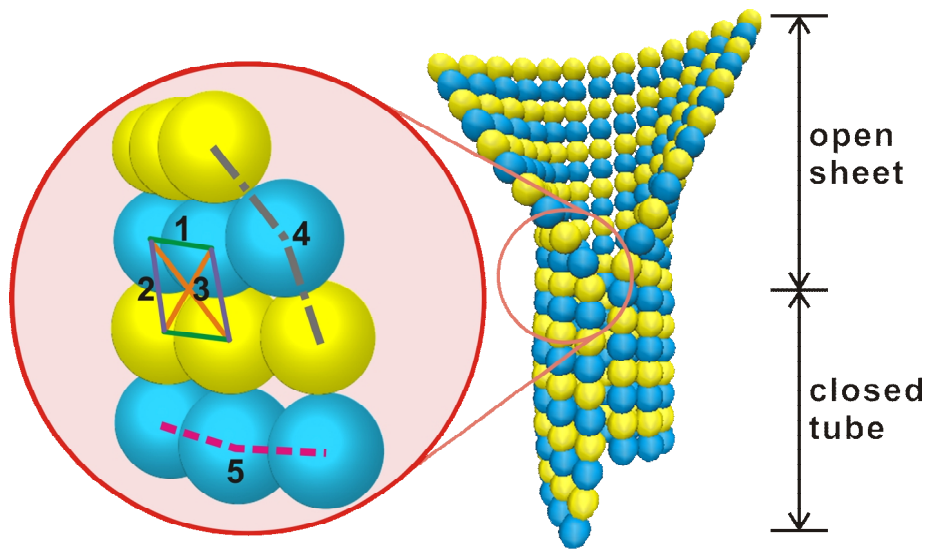

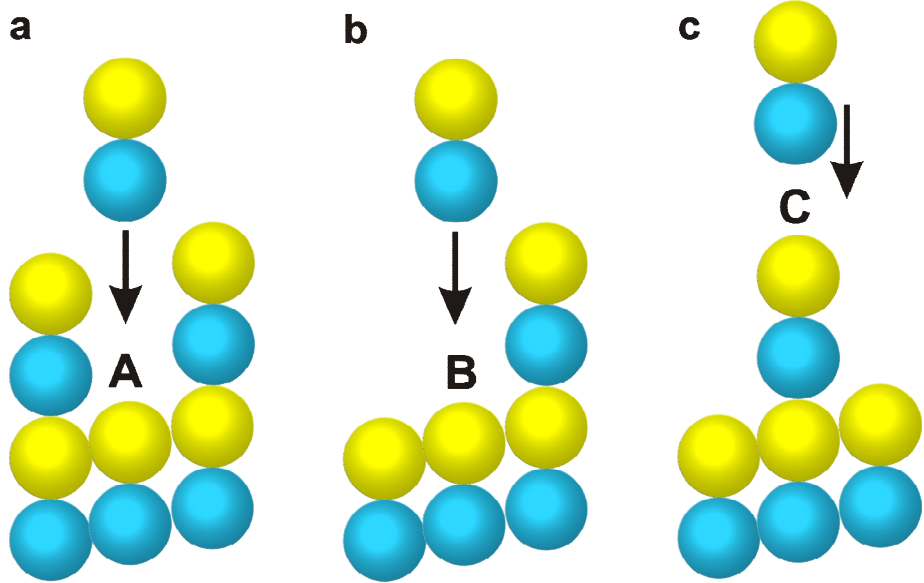

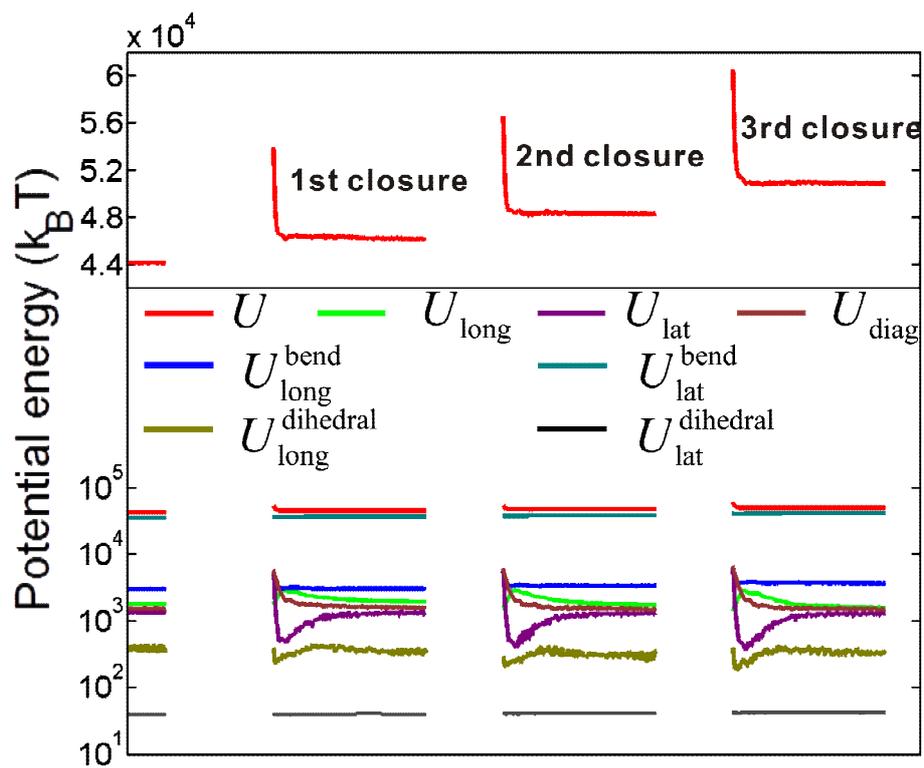

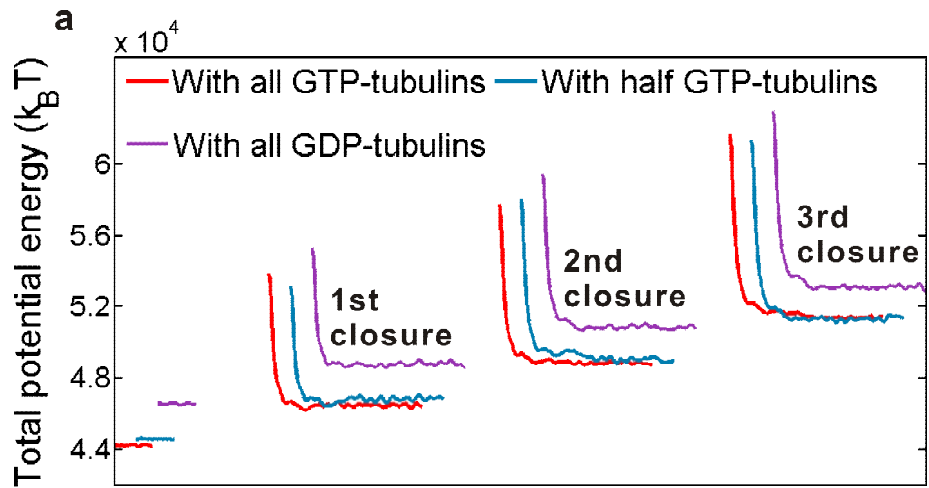

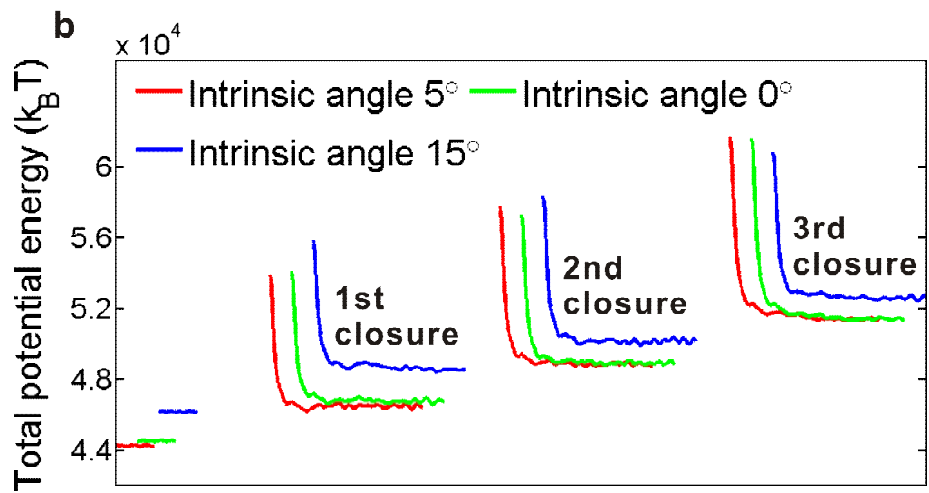

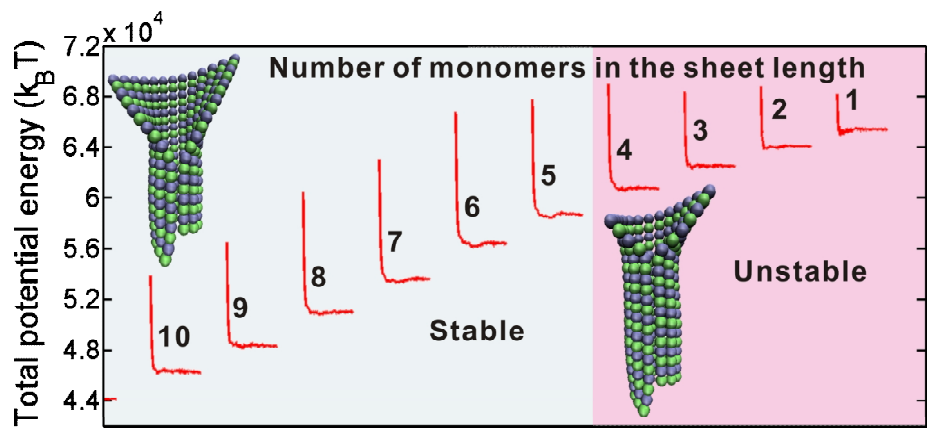

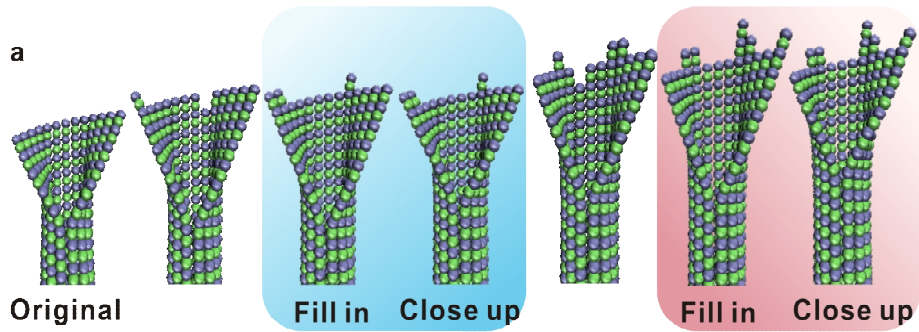

a

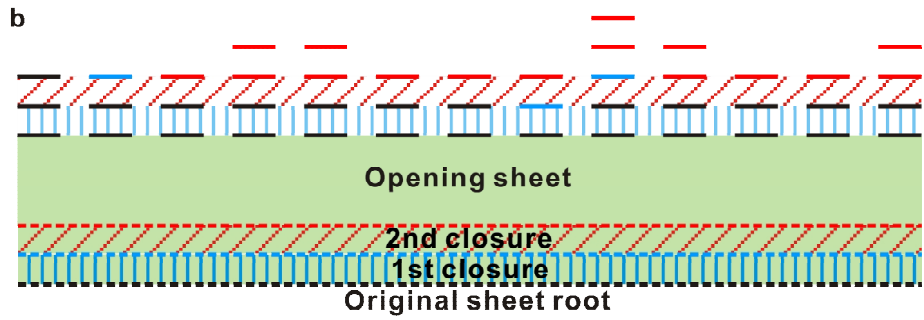

b

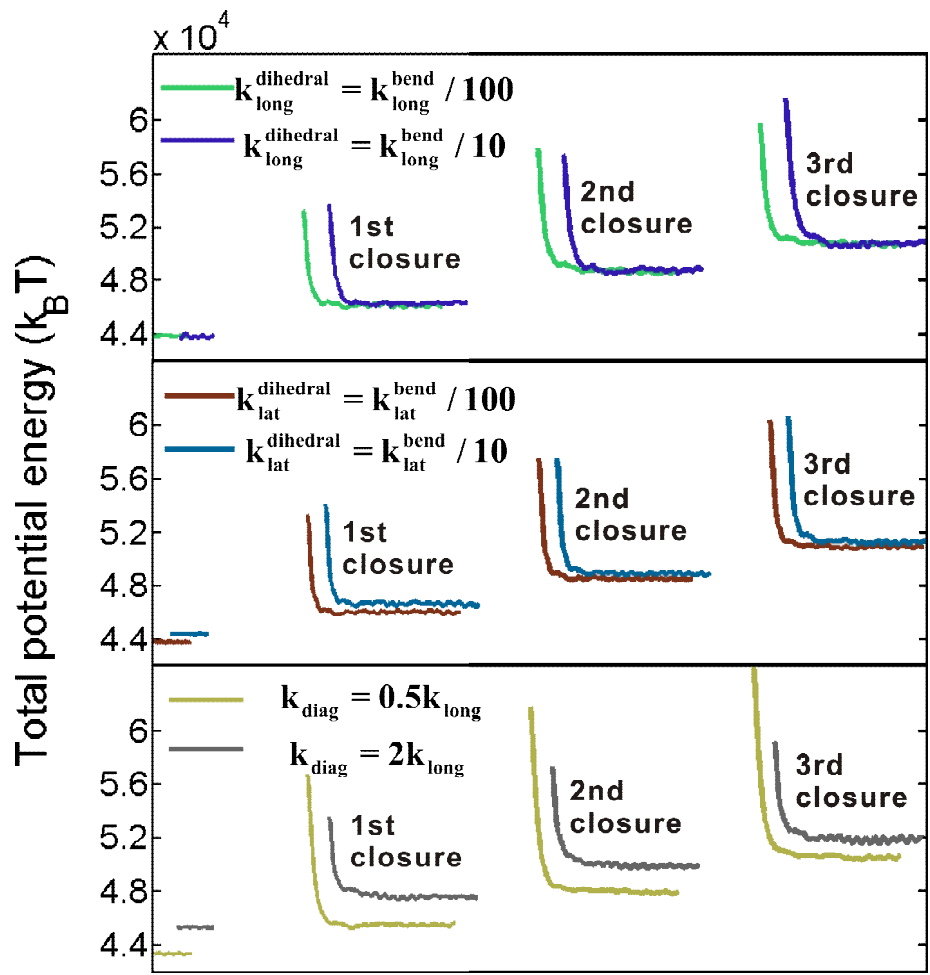